\documentclass[aps,prl,amsmath,amssymb,superscriptaddress,lengthcheck]{revtex4-1}

\usepackage{graphicx}% Include figure files
\usepackage{dcolumn}% Align table columns on decimal point
\usepackage{bm}% bold math
\usepackage{relsize}% for large math symbols
\usepackage{xcolor,color}
\usepackage{ulem}
\usepackage{hyperref}
\hypersetup{colorlinks=true,urlcolor=blue,citecolor=blue}

\definecolor{ehgreen}{rgb}{0, 0.6, 0}

\newcommand{\lnsco}{$\textrm{La}_{1.475}\textrm{Nd}_{0.4}\textrm{Sr}_{0.125}\textrm{Cu}\textrm{O}_{4}$}
\newcommand{\lbco}{$\textrm{La}_{1.885}\textrm{Ba}_{0.115}\textrm{Cu}\textrm{O}_{4}$}
\newcommand{\lbcob}{$\textrm{La}_{1.875}\textrm{Ba}_{0.125}\textrm{Cu}\textrm{O}_{4}$}
\newcommand{\bscco}{$\textrm{Bi}_2\textrm{Sr}_2\textrm{CaCu}_2\textrm{O}_{8+\delta}$}

\newcommand{\ybco}{$\textrm{Y}\textrm{Ba}_2\textrm{Cu}_3\textrm{O}_{6+y}$}
\newcommand{\qco}{$q_{\textrm{CO}}$}
\newcommand{\tltt}{$T_{\textrm{LTT}}$}
\newcommand{\sro}{$\textrm{Sr}_2\textrm{Ru}\textrm{O}_{4}$}

\begin{document}

% Title

\title{Large response of charge stripes to uniaxial stress in \lnsco}

% Place the author information here.  Please hand-code the contact
% information and notecalls; do *not* use \footnote commands.  Let the
% author contact information appear immediately below the author names
% as shown.  We would also prefer that you don't change the type-size
% settings shown here.
\author{T.\,J.\,Boyle}
\email[These authors contributed equally to this work.]{}
\affiliation{Department of Physics, University of California, Davis, California 95616, USA}
\affiliation{Department of Physics, Yale University, New Haven, Connecticut 06520, USA}

\author{M.\,Walker}
\email[These authors contributed equally to this work.]{}
\affiliation{Department of Physics, University of California, Davis, California 95616, USA}

\author{A.\,Ruiz}
\email[These authors contributed equally to this work.]{}
\affiliation{Department of Physics, University of California, San Diego, California 92093, USA}
\affiliation{\mbox{Department of Physics, Massachusetts Institute of Technology, Cambridge, Massachusetts 02139, USA}}

\author{E.\,Schierle}
\affiliation{\mbox{Helmholtz-Zentrum Berlin f\"ur Materialien und Energie, Albert-Einstein-Stra\ss\hspace{0pt}e 15, 12489 Berlin, Germany}}

\author{Z.\,Zhao}
\affiliation{Department of Physics, University of California, Davis, California 95616, USA}

\author{F.\,Boschini}
\affiliation{Quantum Matter Institute, University of British Columbia, Vancouver, British Columbia V6T 1Z4, Canada}
\affiliation{Department of Physics $\&$ Astronomy, University of British Columbia, Vancouver, British Columbia V6T 1Z1, Canada}
\affiliation{Centre \'{E}nergie Mat\'{e}riaux T\'{e}l\'{e}communications, Institut National de la Recherche Scientifique, Varennes, Qu\'{e}bec J3X 1S2, Canada}

\author{R.\,Sutarto}
\affiliation{Canadian Light Source, University of Saskatchewan, Saskatoon, Saskatchewan S7N 2V3, Canada}

\author{T.\,D.\,Boyko}
\affiliation{Canadian Light Source, University of Saskatchewan, Saskatoon, Saskatchewan S7N 2V3, Canada}

\author{W.\,Moore}
\affiliation{Department of Physics, University of California, Davis, California 95616, USA}

\author{N.\,Tamura}
\affiliation{Advanced Light Source, Lawrence Berkeley National Lab, Berkeley, California 94720, USA}

\author{F.\,He}
\affiliation{Canadian Light Source, University of Saskatchewan, Saskatoon, Saskatchewan S7N 2V3, Canada}

\author{E.\,Weschke}
\affiliation{\mbox{Helmholtz-Zentrum Berlin f\"ur Materialien und Energie, Albert-Einstein-Stra\ss\hspace{0pt}e 15, 12489 Berlin, Germany}}

\author{A.\,Gozar}
\affiliation{Department of Physics, Yale University, New Haven, Connecticut 06520, USA}
\affiliation{Energy Sciences Institute, Yale University, West Haven, Connecticut 06516, USA}

\author{W.\,Peng}
\affiliation{Max Planck Institute for Chemical Physics of Solids, N\"othnitzerstrasse 40, 01187 Dresden, Germany}

\author{A.\,C.\,Komarek}
\affiliation{Max Planck Institute for Chemical Physics of Solids, N\"othnitzerstrasse 40, 01187 Dresden, Germany}

\author{A.\,Damascelli}
\affiliation{Quantum Matter Institute, University of British Columbia, Vancouver, British Columbia V6T 1Z4, Canada}
\affiliation{Department of Physics $\&$ Astronomy, University of British Columbia, Vancouver, British Columbia V6T 1Z1, Canada}

\author{C.\,Sch\"u\ss\hspace{0pt}ler-Langeheine}
\affiliation{\mbox{Helmholtz-Zentrum Berlin f\"ur Materialien und Energie, Albert-Einstein-Stra\ss\hspace{0pt}e 15, 12489 Berlin, Germany}}

\author{A.\,Frano}
\email[]{afrano@ucsd.edu}
\affiliation{Department of Physics, University of California, San Diego, California 92093, USA}

\author{E.\,H.\,da Silva Neto}
\email[]{eduardo.dasilvaneto@yale.edu}
\affiliation{Department of Physics, University of California, Davis, California 95616, USA}
\affiliation{Department of Physics, Yale University, New Haven, Connecticut 06520, USA}
\affiliation{Energy Sciences Institute, Yale University, West Haven, Connecticut 06516, USA}

\author{S.\,Blanco-Canosa}
\email[]{sblanco@dipc.org}
\affiliation{Donostia International Physics Center, DIPC, 20018 Donostia-San Sebastian, Basque Country, Spain}
\affiliation{IKERBASQUE, Basque Foundation for Science, 48013 Bilbao, Spain}

%%%%%%%%%%%%%%%%% END OF PREAMBLE %%%%%%%%%%%%%%%%
\begin{abstract}
    The La-based `214' cuprates host several symmetry breaking phases including superconductivity, charge and spin order in the form of stripes, and a structural othorhombic-to-tetragonal phase transition. Therefore, these materials are an ideal system to study the effects of uniaxial stress onto the various correlations that pervade the cuprate phase diagram. We report resonant x-ray scattering experiments on \lnsco~(LNSCO-125) that reveal a significant response of charge stripes to uniaxial tensile-stress of $\sim0.1$\,GPa. These effects include a reduction of the onset temperature of stripes by $\sim50$\,K, a $29$\,K reduction of the low-temperature orthorhombic-to-tetragonal transition, competition between charge order and superconductivity, and a preference for stripes to form along the direction of applied stress. 
    Altogether, we observe a dramatic response of the electronic properties of LNSCO-125 to a modest amount of uniaxial stress.
\end{abstract}

\maketitle 

Cuprate high-temperature superconductors may be the quintessential example of a strongly correlated quantum system, featuring a complex interplay between broken symmetry states, often referred to as intertwined orders \cite{Keimer2015,Tranquada2015}. This rich interplay is evident in the charge order (CO) state: a periodic modulation of charge that is intertwined with superconductivity, magnetism, the crystal structure, and nematicity (four-fold $C_4$ to two-fold $C_2$ symmetry breaking \cite{Fra10}). These various links between ordered states and CO are perhaps most clearly observed in the `214' family of La-based cuprates. \textit{(i)} CO and magnetism emerge intertwined in the form of stripes. \textit{(ii)} The stripes are $C_2$ symmetric, with their orientation alternating by $90^\circ$ between adjacent CuO$_2$ planes \cite{Tran95}. \textit{(iii)} The stripes are strongly pinned to the low temperature tetragonal (LTT) crystal structure, which stabilizes the alternating pattern \cite{Fink09,Hucker11}. \textit{(iv)} Finally, the stabilization of stripes occurs in concert with the suppression of three-dimensional (3D) superconductivity, which results from a frustration of the Josephson coupling between adjacent CuO$_2$ layers \cite{Li07,Tran08}. Determining how all these ordered states are interconnected is a major challenge in the field of high-temperature superconductivity.

\begin{figure*}
	\includegraphics[width=6.75 in]{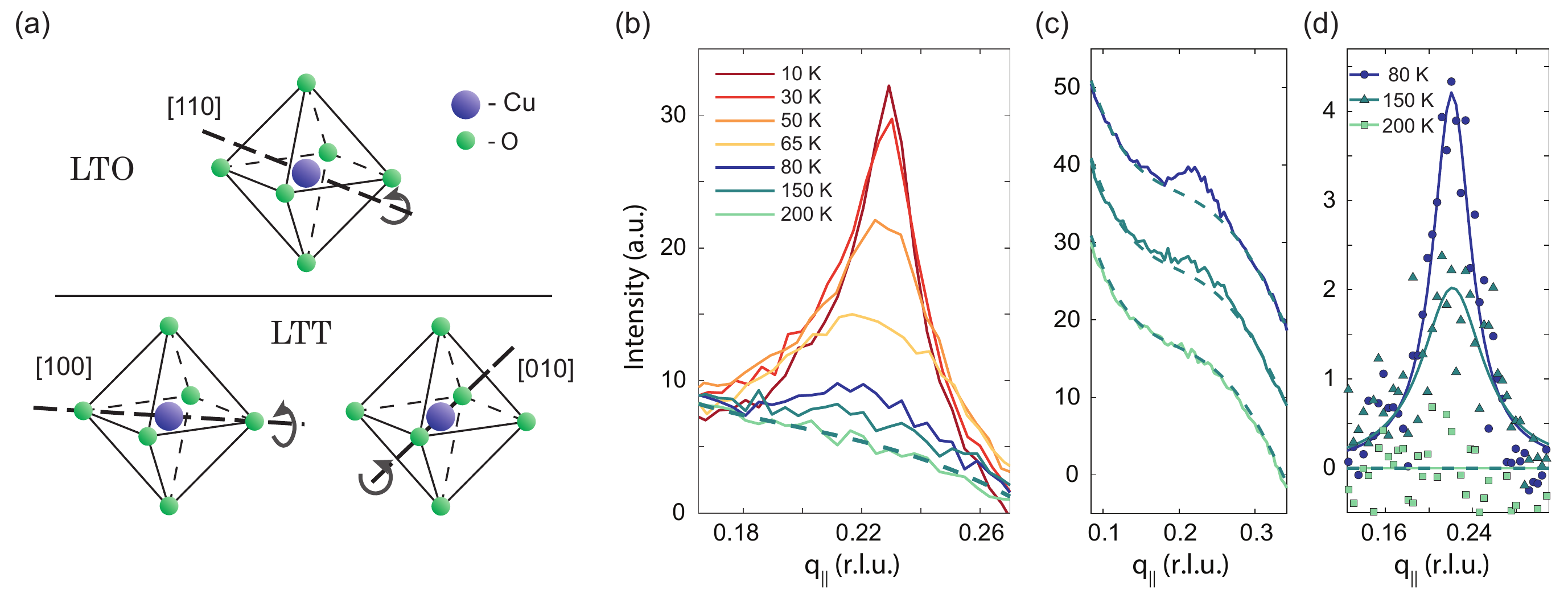}
	\caption{(a) A diagram depicting the CuO$_6$ octahedral tilts in both the LTO and LTT phases. In the LTO phase, neighboring octahedra tilt in opposite directions (clockwise and counter-clockwise) along the [110] direction. In the LTT phase, adjacent CuO$_2$ layers have orthogonal tilts along the [100] and [010] directions. (b) Temperature dependence of the CO peak in LNSCO-125. Sharp peaks corresponding to charge stripes are observed below 65 K in the LTT phase. The gray dashed line is a polynomial fit of the high-temperature background curve measured at 300 K. (c) RXS scans in the LTO phase showing broad peaks corresponding to CO correlations. (d)~Lorentzian fits of the background-subtracted curves shown in (c).}
	\label{fig1}
\end{figure*}

There are different methods to tune the intertwined orders in the cuprates. Typically, a magnetic field \cite{Chang12,Blan13}, chemical doping \cite{Yam98}, or pressure \cite{Cyr18} are used to adjust the relative strengths of the ordered states. For instance, the application of $1.85$\,GPa of hydrostatic pressure to \lbcob~decouples the LTT and CO phases, suppressing the former while allowing the latter to survive with an onset temperature of $35$\,K \cite{Hucker2010}. However, controlling the rotational symmetry of the correlations in the CuO$_2$ planes requires a tuning parameter that couples directionally to the electronic degrees of freedom. This can be achieved with the uniaxial application of pressure, or stress, which has recently become the focus of several cuprate studies. For example, recent experiments on \ybco~indicate that attaining uniaxial strain around 1\% can modify the inter-layer interaction of the CO and its coupling to acoustic phonons \cite{Kim2018,kim_charge_2020}. In the `214' family, uniaxial stress has repeatedly been shown to increase the onset temperature of superconductivity~\cite{Uchida2002,Takeshita2004,Takagi2005,Guguchia2020}. A small value of uniaxial stress, approximately $0.05$\,GPa, dramatically increases the onset of 3D superconductivity in  \lbco~(LBCO-115) from $6$\,K to $32$\,K \cite{Guguchia2020}. Remarkably, the same application of uniaxial stress also reduces the onset of spin stripe order from $38$\,K to $30$\,K. These opposing effects between superconductivity and spin stripes highlight the importance of uniaxial stress experiments to the study of intertwined orders. Still, diffraction experiments that directly measure the effects of uniaxial stress on charge stripe order and the LTT structure in the La-based cuprates have not been reported.

Here we report a Cu-L$_3$ and O-K edge resonant x-ray scattering (RXS) study of CO and the LTT structure in \lnsco~(LNSCO-125) under the influence of modest uniaxial stress, approximately $0.1$\,GPa, applied along the $a$ axis of the LTT structure (\textit{i.e.} along the Cu-O bond direction). We first performed a detailed zero-stress control experiment on LNSCO-125, where we observed CO correlations above the onset of stripes near $70$\,K and detectable up to at least $150$\,K. Upon introducing uniaxial tensile-stress, the onset of the LTT phase (\tltt) is reduced by $29$\,K, from $63$\,K to $34$\,K. Additionally, the onset of stripe order decreases by approximately $50$\,K, from $\approx T_{LTT} + 12$\,K in the absence of strain to $\approx T_{LTT} - 9$\,K in the presence of strain. This overall shift is larger than the one observed for \tltt, which likely reflects the competition between CO and superconductivity. An additional offset of approximately $6$\,K is observed between the onset of charge stripes along the $a$ and $b$ directions. Despite the effects of stress on the LTT and the charge stripe phases, we find no appreciable modification of the high-temperature CO correlations. Altogether, our experiments not only show that a small amount of uniaxial stress triggers responses from the various intertwined orders, they also establish uniaxial stress as a powerful tool to control the electronic properties of LNSCO-125.

The single crystals were synthesized by floating zone and previously characterized by means of resonant x-ray scattering and magnetometry \cite{Blan18}. LNSCO-125 hosts several classic features of La-based cuprates near $1/8$ hole doping. In the low temperature orthorhombic (LTO) phase, the CuO$_6$ octahedra tilt about an axis parallel to the $[110]$ direction, which is $45^\circ$ from the Cu-O bond direction. At lower temperatures, in the LTT phase, the CuO$_6$ octahedra tilt about axes parallel the Cu-O bond directions \cite{Fujita1989, Xu1989}. Importantly, in the LTT phase the tilt axis alternates between $[100]$ and $[010]$ through consecutive CuO$_2$ planes, Fig.\,\ref{fig1}(a). Thus, while the LTT is a globally tetragonal phase, it is actually two-fold $C_2$ symmetric within each CuO$_2$ plane, which provides a natural motif to stabilize the stripe order. 

\begin{figure}
	\includegraphics[width=\linewidth]{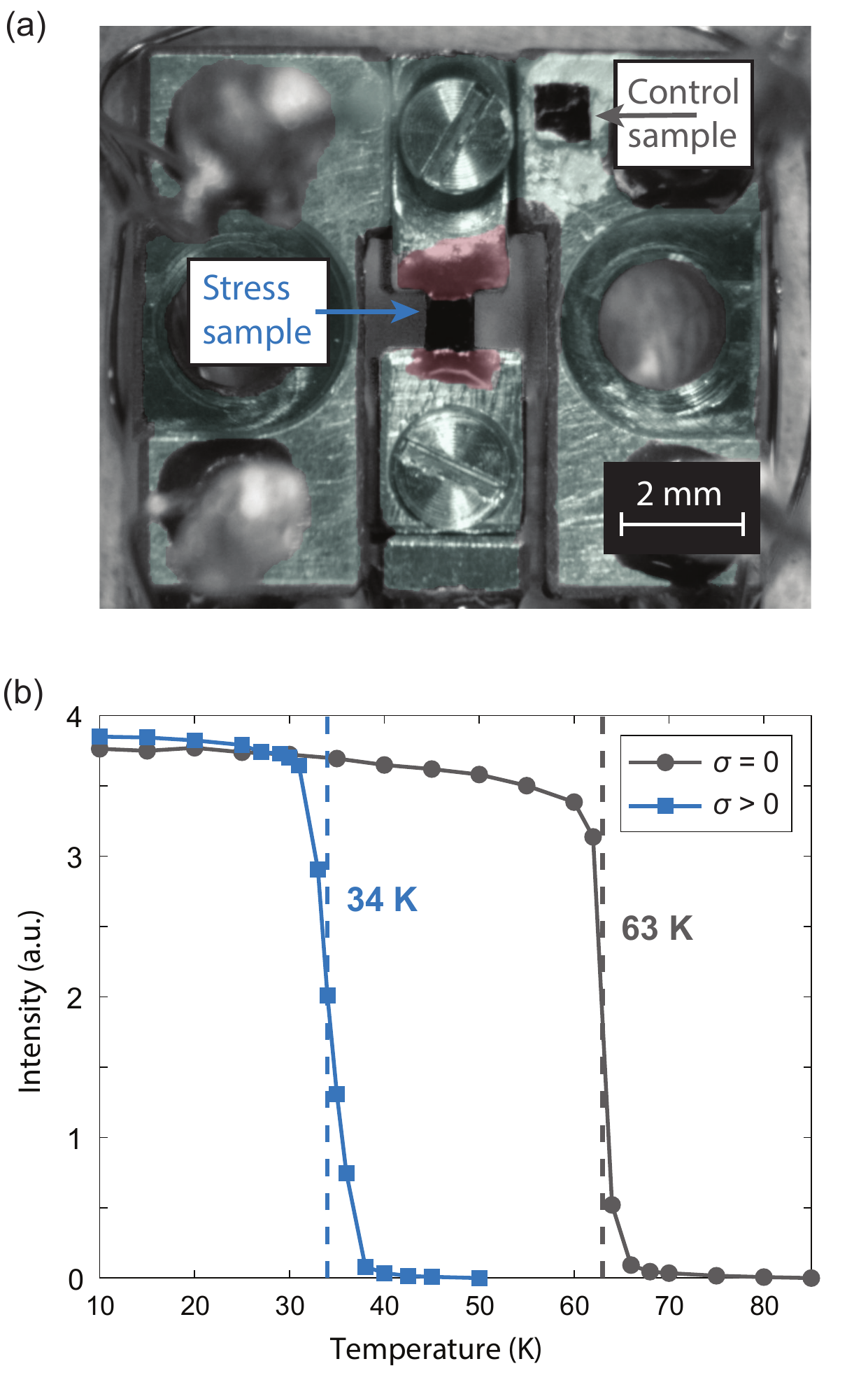}
	\caption{(a) An image of the titanium strain device and LNSCO-125 crystals. The stress sample is mounted in the center across a gap using a stiff epoxy and the control sample is mounted on the titanium using silver paint \cite{SM}. (b) Temperature dependence of the (001) Bragg peak on the apical O-K edge in both the stress and control samples corresponding to the LTT phase transition.  The transition temperature is suppressed by 29 K with the applied stress.}
	\label{fig2}
\end{figure}

Previous RXS experiments on LNSCO-125 indicate the appearance of stripe order at approximately \tltt\,\cite{Achkar2016}.  Our detailed RXS experiments show that the peak at $q_{||} = q_{CO} = 0.23$\,rlu (reciprocal lattice units) survives at high temperatures, above \tltt, albeit with much lower intensity when compared to the low-temperature signal. Figure \ref{fig1}(b) shows the temperature evolution of the CO peak at \qco~in LNSCO-125, showing a clear and rapid enhancement below $T_{LTT} = 63$\,K. However, the data also show that correlations at \qco~persist even for $T>T_{LTT}$ and continue to decrease with increasing temperature. It is difficult to determine the onset temperature for these correlations, but we still observe a clear evolution of the peak at \qco~between $80$\,K and $200$\,K, Fig.\,\ref{fig1}(c), which is better visualized by subtracting the $300$\,K curve, Fig.\,\ref{fig1}(d).

It is important to note that since our RXS experiments are done in energy-integrated mode, the high-temperature peak at \qco~may originate from both elastic (static) and inelastic (dynamic) correlations. In fact, high-temperature dynamic CO signals have recently been observed in many cuprates, including other La-based systems \cite{Dean2019,Ghiringhelli2019}. While it may be tempting to assign the same rotational symmetry of the charge stripes to the high-temperature correlations, such correspondence has not been experimentally verified. Here we refer to the low temperature signal as charge stripes and cautiously refer to the high-temperature signal simply as CO correlations. As we will discuss later, we do not detect modifications to the CO correlations due to uniaxial stress in our experiments despite observing significant effects to the LTT phase and charge stripes.  

To investigate what happens to the CO and the LTT phase when we perturb the LNSCO-125 sample with extrinsic uniaxial stress, $\sigma\neq0$, we embed the crystal in an apparatus whose geometry explicitly breaks $C_4$ symmetry, Fig.\,\ref{fig2}(a). The sample is constrained on two-edges across a gap using expoy in a device constructed of machined high-purity titanium \cite{SM}. Differential thermal contraction occurs upon cooling due to the different coefficients of the thermal expansion of the sample, epoxy and titanium, which causes the LNSCO-125 crystal to be uniaxially stretched relative to an unconstrained crystal \cite{SM}. Our experimental setup also includes a second sample mounted directly on one of the faces of the device using silver paint, which allows us to perform control measurements on a $\sigma=0$ sample in the same experiment, Fig.\,\ref{fig2}(a). (Note that $\sigma$ refers to externally applied stress and does not include the intrinsic stress due to the thermal contraction of the crystals themselves.) Unlike many strain experiments that cannot directly probe the lattice parameters, we can access the lattice constants by measuring the Bragg peaks of the  LNSCO-125 crystal. We find those measurements to yield strain values of $\approx 0.046\pm 0.026\,\%$ \cite{SM}. Using $C_{11} = 232$\,GPa for the elastic modulus \cite{Fujita1995}, we estimate $\sigma = 0.11\pm0.06$\,GPa. Additionally, we perform a multiphysics simulation that incorporates all key elements of the assembly, including the thermal and elastic properties of the materials, as well as the geometry of the assembly \cite{SM,chang2016}. The simulation produces an approximately uniform tensile strain pattern of the same order of magnitude as measured by the Bragg peaks when the apparatus is cooled to $70$\,K. Although the amount of stress is relatively small, it is comparable in magnitude to the compressive stress used in the LBCO-115 experiments mentioned above, which were shown to have significant effects on the superconductivity and spin-stripe order \cite{Guguchia2020}.

A striking consequence of $\sigma\neq0$ in our experiments on LNSCO-125 is the dramatic reduction of \tltt~from $63$\,K to $34$\,K, Fig.\,\ref{fig2}(b). This is directly seen in our experiments via apical O-K edge RXS measurements of the (001) Bragg peak (in Miller index notation), whose resonant cross-section is an increasing function of the octahedral tilts in the LTT phase \cite{Achkar2016}. We can understand the reduction of \tltt~as a  consequence of uniaxial stress, which spoils the global tetragonality of the LTT phase. Although it still emerges at low temperatures, our experiments unveil a remarkable response of the LTT structure in LNSCO-125 to the application of uniaxial stress. 

In principle, uniaxial stress could result in the transition of the macroscopic crystal into a mixed LTO/LTT phase or a decrease in the LTT octahedral tilt angle, as hypothesized from measurements of the magnetic properties of LBCO-115 \cite{Guguchia2020}; neither effect is resolved in our experiments. First note that the intensity of the (001) Bragg peak, whose cross-section is an increasing function of the octahedral tilt angle, appears unchanged by $\sigma$ at 10 K. This indicates that, within the LTT phase, the structure is unaffected by stress. Second, the LTO-LTT transition remains sharp with non-zero $\sigma$, which suggests that the stressed sample enters the LTT phase in a rather uniform fashion. While the uniformity of the strain field on the sample may vary between different uniaxial strain setups, our experiments do not provide evidence of a mixed phase or reduced octahedral tilt angle.

\begin{figure}
	\includegraphics[width=\linewidth]{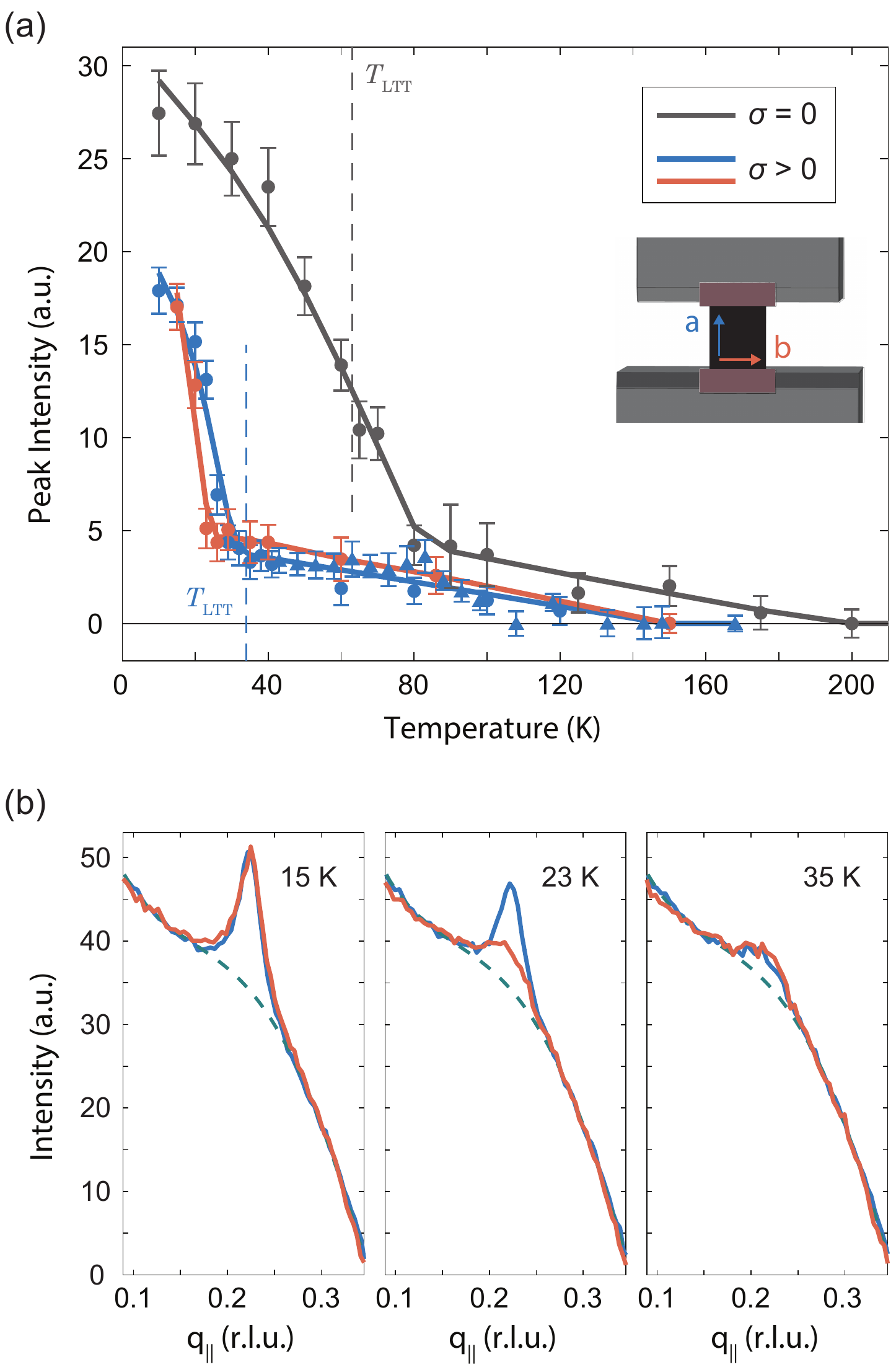}
	\caption{(a) Temperature dependence of the CO peaks in the stress and control samples. The solid lines serve as guides to the eye. The dashed lines correspond to the LTT transition temperatures determined from Fig.\,\ref{fig2}(b). The inset shows a diagram of the stress sample, where we label $a$ (blue) and $b$ (orange) as the directions parallel to and perpendicular to the direction applied stress respectively. Measurements of the stress sample along the $a$ direction were performed at two different synchrotrons: measurements from the Canadian Light Source are triangles and measurements from BESSY-II are circles \cite{SM}. (b) Comparison of stress sample RXS scans along the $a$ and $b$ directions near the LTT transition. The gray dashed line is a polynomial fit of the high-temperature background curve measured at 150 K. The onset of the charge stripes along the direction of applied stress precedes that of the perpendicular direction by approximately $6$\,K.}
	\label{fig3}
\end{figure}

Within the same experiment we can also track the temperature dependence of the charge order in the stressed LNSCO-125 crystal, which is summarized in Fig.\,\ref{fig3}(a). A consequence of uniaxial stress is the reduction of the onset temperature of charge stripes by $50$\,K, which is larger than the change of $29$\,K observed for \tltt. If the only effect of uniaxial stress was the suppression of \tltt, one might have expected the same change to occur for the onset of charge stripes. However, the additional shift of $21$\,K suggests that one should also consider interactions with additional intertwined orders, such as superconductivity. Uniaxial stress on the order of $0.05$\, GPa has been shown to increase $T_c$ by $8$\,K in LNSCO-120 and by $10$\,K in LBCO-115 \cite{Uchida2002,Guguchia2020}, showing that superconductivity is enhanced in tandem with the suppression of the LTT phase. Additionally, cuprates that lack a similar structural transition, such as \bscco~and \ybco, exhibit competition between superconductivity and CO \cite{Ghiringhelli2012,Chang12,Silva-Neto2014}. Together, the suppression of the LTT phase and the enhancement of superconductivity would account for the larger suppression of the onset of stripes, relative to the suppression of \tltt.

In addition to the effects on stripe order due to its intertwining with the LTT phase and superconductivity, uniaxial stress may also directly influence the pinning of stripes along $a$ and $b$. For example, given the $C_2$ symmetry of stripe order, one may expect that a finite $\sigma$ in the absence of an LTT phase would cause the onset temperatures of stripes along $a$ and $b$ to split. Indeed, this split is resolved by our detailed temperature dependent RXS measurements, Fig.\,\ref{fig3}(a). This is seen more clearly by directly comparing the RXS data along the two directions, Figure \ref{fig3}(b), which shows that at $23$\,K the peak along $a$ (parallel to the applied stress) has already entered the charge stripe phase, while the peak along $b$ is still very similar to the peak in the high-temperature phase -- compare to the $35$\,K data. Eventually, at $15$\,K the two signals approach the same saturation value, which may indicate that the LTT structure has suppressed the effects of $\sigma \approx 0.1$\,GPa at this temperature. Nevertheless, our observations show that uniaxial stress can be used to pin the direction of stripes in the CuO$_2$ plane.

While our measurements show a delicate balance between charge stripes, superconductivity and structural distortions, a direct effect of uniaxial stress on the high-temperature CO correlations is not clearly observed in our data. Just above the onset of charge stripe order, the intensities of the RXS peaks at \qco~are indistinguishable between the stressed and control samples. We also do not resolve a clear difference between the CO correlations along $a$ and $b$ in the stressed sample -- for example see the $35$\,K data in Fig.\,\ref{fig3}(b). Figure \ref{fig3}(a) suggests that the onset of CO correlations at high-temperatures may be impacted by the application of stress along $a$. However, there are several complications that prevent us from reaching that conclusion with any reasonable confidence. First, as mentioned above, the CO correlations evolve very slowly with temperature, which makes the assignment of an onset temperature difficult. Second, the stress produced by our apparatus is temperature dependent and we estimate a 3 to 5 fold decrease in the applied stress from $34$\,K to $200$\,K. Third, the comparison of small RXS peaks between the stress and control samples can be influenced by variations in the fluorescence background, which is sensitive to sample surface conditions. The effects of uniaxial stress to the high-temperature correlations will likely require experiments that tune the stress at a fixed temperature. Altogether, we cannot conclude the observation of any direct coupling between uniaxial stress and high-temperature CO correlations.     

Our experiments demonstrate the complex relationship between uniaxial stress, charge stripes and the low-temperature tetragonal structure in LNSCO-125. Increasing $\sigma$ from zero to $0.1$\,GPa causes a simultaneous reduction in the onset temperatures of both the LTT phase and charge stripes, as well as a temperature splitting in the formation of charge stripes along the \textit{a} and \textit{b} directions. Additionally, the effects of superconductivity need to be included to describe our observations, with the competition between superconductivity and charge order together with the enhancement of $T_c$ under uniaxial stress serving as a natural explanation for the additional suppression of the onset of stripe order with respect to \tltt. Furthermore, we find that larger stress may be necessary to cause significant changes to the high-temperature CO correlations. Nevertheless, the relatively small amount of stress necessary to tune the electronic properties of the La-based cuprates near $1/8$ hole-doping is quite remarkable. For example, strain on the order of $1.0$\,\% is necessary to modify the properties of charge order in \ybco~\cite{kim_charge_2020} or to shift the superconducting transition by $2$\,K in \sro ~\cite{Hicks14,Step17}. While achieving $1.0$\,\% strain may be quite challenging and difficult to reproduce, $0.05$\,\% or even smaller is clearly sufficient to alter the electronic properties of LNSCO-125, which opens new opportunities for switchable devices and precision detectors at the current frontier of technology. 

\begin{acknowledgments}
The research described in this paper was carried out at UE-46-PGM1 and UE-56/2-PGM2 beamlines of BESSY II, a member of the Helmholtz Association (HGF), and at the Canadian Light Source, a national research facility of the University of Saskatchewan, which is supported by the Canada Foundation for Innovation (CFI), the Natural Sciences and Engineering Research Council (NSERC), the National Research Council (NRC), the Canadian Institutes of Health Research (CIHR), the Government of Saskatchewan, and the University of Saskatchewan. This material is based upon work supported by the National Science Foundation under Grant No.\,1845994 and 2034345. Part of the research leading to these results has been supported by the project CALIPSO and under the Grant Agreement 730872 from the EU Framework Programme for Research and Innovation HORIZON 2020. S.B.-C. acknowledges the MINECO of Spain for financial support through the project PGC2018-101334-A-C22. A. R. acknowledges support from the University of California President's Postdoctoral Fellowship Program. This work was supported by the Alfred P. Sloan Fellowship (E.H.d.S.N. and A.F.). The research in Dresden is supported by the Deutsche Forschungsgemeinschaft through Grant No. 320571839. This research was undertaken thanks in part to funding from the Max Planck-UBC-UTokyo Centre for Quantum Materials and the Canada First Research Excellence Fund, Quantum Materials and Future Technologies Program.
\end{acknowledgments}

\bibliographystyle{apsrev4-1}

\end{document}